\documentclass[sigconf]{acmart}

\AtBeginDocument{%
 }
  
\usepackage{cite}
\usepackage{amsmath,amsfonts} 
\usepackage{textcomp}
\usepackage{xcolor}
\usepackage{booktabs}
\usepackage{multirow,tabularx}
\usepackage{wrapfig}
\usepackage{epsfig}
\usepackage{makecell}
\usepackage{subfigure}
\usepackage{graphicx}
\usepackage{utfsym}
\usepackage{hyperref}
\usepackage{url}
\usepackage{tablefootnote}
\usepackage{caption}
\usepackage{fancyhdr}

\makeatletter
\newcommand\figcaption{\def\@captype{figure}\caption}
\newcommand\tabcaption{\def\@captype{table}\caption}
\makeatother

\copyrightyear{2023}
\acmYear{2023}
\setcopyright{rightsretained}
\acmConference[SIGIR '23]{Proceedings of the 46th International ACM SIGIR Conference on Research and Development in Information Retrieval}{July 23--27, 2023}{Taipei, Taiwan}
\acmBooktitle{Proceedings of the 46th International ACM SIGIR Conference on Research and Development in Information Retrieval (SIGIR '23), July 23--27, 2023, Taipei, Taiwan}\acmDOI{10.1145/3539618.3591932}
\acmISBN{978-1-4503-9408-6/23/07}

\makeatletter
\gdef\@copyrightpermission{
  \begin{minipage}{0.3\columnwidth}
  \href{https://creativecommons.org/licenses/by/4.0/}{\includegraphics[width=0.90\textwidth]{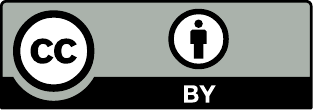}}  
\end{minipage}\hfill
  \begin{minipage}{0.7\columnwidth}
  \href{https://creativecommons.org/licenses/by/4.0/}{This work is licensed under a Creative Commons Attribution International 4.0 License.}
  \end{minipage}
  \vspace{5pt}
}
\makeatother

\begin{document}

\title{Where to Go Next for Recommender Systems? ID- vs. Modality-based Recommender Models Revisited}

\author{Zheng Yuan$^{*}$}
\affiliation{
  \institution{Westlake University}\streetaddress{}\city{}\country{}}
\email{yuanzheng@westlake.edu.cn}

\author{Fajie Yuan$^{*\dagger}$}
\affiliation{
  \institution{Westlake University}\streetaddress{}\city{}\country{}}
\email{yuanfajie@westlake.edu.cn}

\author{Yu Song}
\affiliation{
  \institution{Westlake University}\streetaddress{}\city{}\country{}}
\email{songyu@westlake.edu.cn}

\author{Youhua Li}
\affiliation{
  \institution{Westlake University}\streetaddress{}\city{}\country{}}
\email{liyouhua@westlake.edu.cn}  

\author{Junchen Fu}
\affiliation{
  \institution{Westlake University}\streetaddress{}\city{}\country{}}
\email{fujunchen@westlake.edu.cn}

\author{Fei Yang}
\affiliation{
  \institution{Zhejiang Lab}\streetaddress{}\city{}\country{}}
\email{yangf@zhejianglab.com}

\author{Yunzhu Pan}
\affiliation{
  \institution{Westlake University}\streetaddress{}\city{}\country{}}
\email{panyunzhu@westlake.edu.cn} 

\author{Yongxin Ni}
\affiliation{
  \institution{Westlake University}\streetaddress{}\city{}\country{}}
\email{niyongxin@westlake.edu.cn}

\thanks{$*$ Equal Contribution.}\thanks{$\dagger$ Corresponding author. Fajie designed and supervised this research; Zheng performed this research, in charge of key technical parts;  Fajie, Zheng, Yu wrote the manuscript. Yunzhu collected the Bili dataset, other authors assisted partial experiments.}
\renewcommand{\shortauthors}{Zheng Yuan et al.}

\begin{abstract}
  
Recommendation models that utilize unique identities (IDs for short) to represent distinct users and items have been state-of-the-art (SOTA) and dominated the recommender systems (RS) literature for over a decade. Meanwhile, the pre-trained modality encoders, such as BERT~\citep{devlin2018bert} and Vision Transformer~\citep{dosovitskiy2020image}, have become increasingly powerful in modeling the raw modality features of an item, such as text and images. Given this, a natural question arises: can a purely modality-based recommendation model (MoRec) outperforms or matches a pure ID-based model (IDRec) by replacing the itemID embedding with a SOTA modality encoder? In fact, this question was answered ten years ago when IDRec beats MoRec by a strong margin in both recommendation accuracy and efficiency.

We aim to revisit this `old' question and systematically study MoRec from several aspects. Specifically, we study several sub-questions: (i) which recommendation paradigm, MoRec or IDRec, performs better in practical scenarios, especially in the general setting and warm item scenarios where IDRec has a strong advantage? does this hold for items with different modality features? (ii) can the latest technical advances from other communities (i.e., natural language processing and computer vision) translate into accuracy improvement for MoRec? (iii) how to effectively utilize item modality representation, can we use it directly or do we have to adjust it with new data? (iv) are there any key challenges that MoRec needs to address in practical applications? To answer them, we conduct rigorous experiments for item recommendations with two popular modalities, i.e., text and vision. We provide the first empirical evidence that MoRec is already comparable to its IDRec counterpart with an expensive end-to-end training method, even for warm item recommendation. Our results potentially imply that the dominance of IDRec in the RS field may be greatly challenged in the future. We release our code and other materials at \textcolor{blue}{https://github.com/westlake-repl/IDvs.MoRec}.
\end{abstract}

\ccsdesc[500]{Information systems~Recommender systems}

\keywords{Recommender Systems, ID-based Recommendation, Modality-based Recommendation, End-to-end Training}

\maketitle

   \section{Introduction}\label{section:Introduction}
    Recommender system (RS)  models learn the historical interactions of users and items and recommend items that users may interact with in the future. RS is playing a key role in search engines, advertising systems, e-commerce websites, video and music streaming services, and various other Internet platforms.  
    The modern recommendation models usually use unique identities (ID) to represent users and items, which  are subsequently converted to embedding vectors as learnable parameters.
    These ID-based recommendation models (IDRec) have been well-established and dominated the RS field for over a decade until now~\citep{koren2009matrix,rendle2012bpr,yuan2022tenrec}.
 
    Despite that, IDRec  has key weaknesses that can not be ignored. First,
    IDRec highly relies on the ID interactions, which fails to provide recommendations when users and items have few interactions~\citep{yuan2020parameter,yuan2021one}, a.k.a. the cold-start setting. Second, pre-trained IDRec is not transferable across platforms given that userIDs and itemIDs are in general not shareable in practice. This issue seriously limits the development of big \& general-purpose  RS models~\citep{ding2021zero,bommasani2021opportunities,wang2022transrec}, an emerging paradigm in the deep learning  community.
   Third, pure IDRec 
    cannot benefit from technical advances in other communities, such as powerful foundation models (FM)~\citep{bommasani2021opportunities} developed in 
    NLP (natural language processing) and CV (computer vision) areas.
   Moreover, maintaining a large and frequently updated ID embedding matrix for users and items remains a key challenge in industrial applications~\citep{sun2020generic}.  Last but not the least, recommendation models leveraging ID features have obvious drawbacks in terms of interpretability, visualization and evaluation.

One way to address these issues is to replace the ID embedding (of IDRec) with an item modality encoder (ME), especially when item modality  features such as images and text are available.
We refer to such recommendation models as MoRec. In fact, such MoRec appeared in literature many years ago but it was mainly used to solve cold-start or cross-domain recommendation problems~\citep{van2013deep,fu2019deeply,chen2017fully}. In other words, MoRec is rarely adopted when recommending non-cold or popular items unless combined with other effective features, such as the itemID features, e.g., in~\citep{he2016vbpr,kim2016convolutional,wang2011collaborative}.
A key reason might be that these item ME developed in the past years (e.g., word embedding~\citep{mikolov2013distributed} and some shallow   neural networks~\citep{van2013deep}) are not as expressive as typical itemID embeddings.
Today, however, given the recent great  success of FM, we think it is time to revisit the key comparison between modern  MoRec  and IDRec, especially for regular (or non cold-item) recommendation. For example,
    BERT~\citep{devlin2018bert}, GPT-3~\citep{brown2020language} and various Vision Transformers (ViT)~\citep{dosovitskiy2020image,liu2021swin} have revolutionized the NLP and CV fields when representing  textual and visual features. Whether item representations learned by them are better suited for the \textit{regular}  recommendation task than ID features remains unknown.
    
    In this paper, we intend to rethink the potential of MoRec and investigate a key question: \textit{should we  still stick to the IDRec paradigm for future recommender systems?} We concentrate on item recommendation based on the text and vision  modalities --- the two most common modalities in literature.
    To be concise,  we attempt to  address the following  sub-questions:
    
   \textbf{Q(\romannumeral1):
   Equipped  with strong modality encoders (ME),
   can MoRec be comparable to or even surpass IDRec in regular, especially in warm-start item recommendation scenario?}
   To answer this question, we conduct empirical studies by taking into account the  \textbf{two} most representative recommendation architectures (i.e., two-tower based DSSM~\citep{huang2013learning,rendle2020neural} and session-based SASRec~\citep{kang2018self}) equipped  with \textbf{four} powerful ME
   evaluated on \textbf{three} large-scale recommendation datasets with \textbf{two}  modalities (text and vision). 

\textit{Novelty clarification}: Though much previous literature has studied MoRec and compared with many baselines~\citep{wu2019neural,ni2019justifying,li2022miner,zhang2021unbert,wu2021empowering}, unfortunately none of them provided a \textit{fair} or rigorous comparison between their proposed MoRec and the corresponding IDRec counterparts in  regular or even  warm item recommendation setting. Fair comparison here means that MoRec and IDRec should 
at least be compared with the same \textit{backbone network} and \textit{experimental settings}, such as samplers and loss functions.
Without a fair comparison,
the community  can not  truly assess the progress of MoRec and the expressive power of ME for recommendation.
   
   \textbf{Q(\romannumeral2):
   If Q(i) is yes, can the recent technical advances developed in NLP and CV fields translate into accuracy improvement in MoRec when using text and visual  features? }
We address this question by performing three experiments. First,
we evaluate MoRec by comparing smaller \emph{vs} larger ME given that pre-trained ME with larger model sizes tends to perform better than their smaller counterparts in various downstream tasks; second, we evaluate MoRec by comparing weaker \emph{vs} stronger ME where weaker and stronger are determined by NLP and CV tasks;
   third,  we evaluate MoRec
   by comparing ME 
   with \emph{vs} without pre-training on corresponding NLP and CV datasets.

   \textbf{Q(\romannumeral3):
Are the representations learned by these foundation models as general as claimed? 
   How can we effectively use item modality representations derived from an NLP or CV encoder network? 
   }
A desirable goal of FM research is to develop models that generate universal representations that can be directly used for various downstream tasks~\citep{lin2022could}. We examine this by first extracting frozen modality features from  well-known ME and then adding them as common features for recommendation models, often referred to as the two-stage (TS) paradigm. This is a common practice for large-scale industrial recommender systems due to training efficiency consideration~\citep{mcauley2015image,covington2016deep}. We then compare TS  with  joint or end-to-end (E2E) training of both the recommendation architecture and ME.

\textit{Novelty clarification}: Though several recent literature has explored E2E learning~\citep{wu2021empowering,wu2021newsbert,yang2022gram,xiao2022training} for recommendation, few of them explicitly discussed the substantial accuracy  and efficiency gap (more than 100x) between TS and E2E paradigms. More importantly, most of them only discussed the DSSM architecture (or other two-tower variants)  without considering  more powerful and computationally more expensive sequence-to-sequence (seq2seq) training approach (e.g., used in SASRec and NextItNet~\citep{yuan2019simple}). 
Furthermore, all of them are only for text recommendation, and so far there is no \textit{modern} (last 5 years) \textit{peer-reviewed} literature considering the E2E learning paradigm for image recommendation.

 
  
 In addition to the aforementioned key questions, we have also identified several challenges that remain unexplored for MoRec when utilizing the end-to-end learning paradigm.


    \section{IDRec \&  MoRec}

    \begin{figure*}[t] 
      \centering
      \includegraphics[height=1.6in]{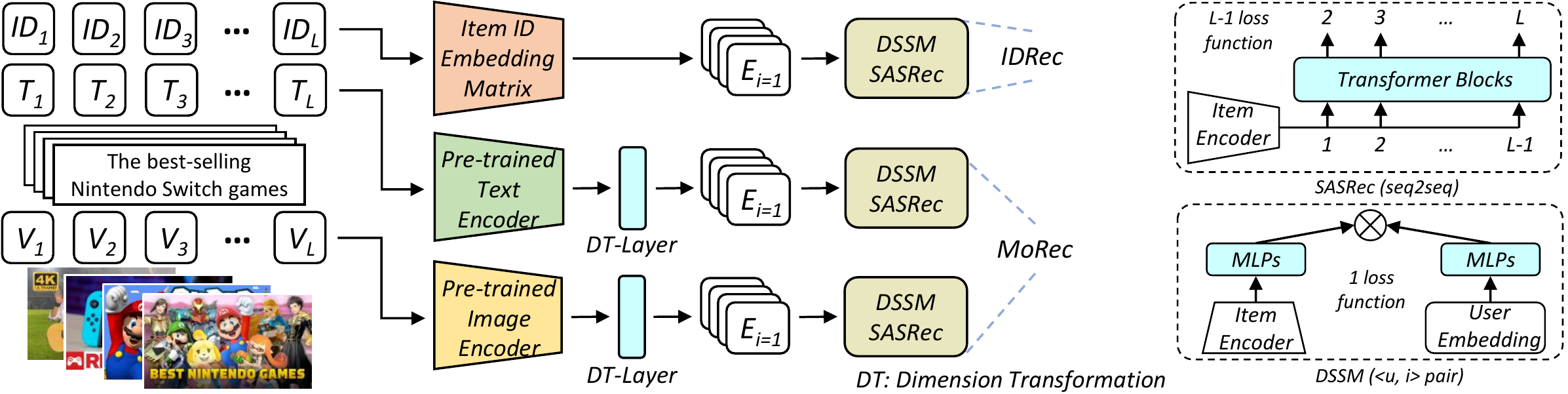}
      \caption{Illustration of IDRec \emph{vs} MoRec. $V_i$ and $T_i$ denote raw features of vision and text modalities. $E_i$ is the  item representation vector fed into the recommender model.
			The only difference between IDRec and MoRec is the item encoder.
			IDRec uses an itemID embedding 
   matrix as the item encoder, whereas MoRec uses the pre-trained ME (followed by a dense layer for the dimension transformation, denoted by DT-layer) as the item encoder.}
      \label{fig:IDRec_and_MoRec}
    \end{figure*}
  
   One core function of a recommendation model is to represent items and users and calculate their matching score. Denote $\mathcal{I}$ (of size $|\mathcal{I}|$) and $\mathcal{U}$ (of size $|\mathcal{U}|$)  as the set of items and users, respectively. For an item $i \in \mathcal{I}$, we can represent it either by its unique ID $i$ or its modality content, such as text and visual features. Likewise, for a user $u \in \mathcal{U}$, we can represent her either by  the unique ID $u$ or the profile of  $u$, where a profile can be the demographic information or 
   a sequence of interacted items. 
   
   In IDRec, an ID embedding matrix $\textbf{X}^{\mathcal{I}} \in \mathbb{R}^{|\mathcal{I}|\times d}$ is initialized, where $d$ is the embedding size. Each vector in  $\textbf{X}^{\mathcal{I}}$ represents the latent space of an item $i$, and can be viewed as a simple item encoder.
   During training and inference, IDRec retrieves ${\textbf{X}^{\mathcal{I}}}_{i} \in \mathbb{R}^{d}$ from $\textbf{X}^{\mathcal{I}}$ as the embedding of $i$ and then feeds it to the recommendation network.
   
   In MoRec, items are assumed to contain modality information. 
   For item $i$, MoRec uses ME to generate the representation for the raw modality feature of $i$ and uses it to replace the ID embedding vector in IDRec. For instance, in the news recommendation scenario, we can use the pre-trained  $\text{BERT}$  or $\text{RoBERTa}$~\citep{liu2019roberta} as text ME and represent a piece of news by the output textual representation of its title. Similarly, when items contain visual features,  we can simply use a pre-trained ResNet or ViT as vision ME. 
   
In this paper, we perform rigorous empirical studies on two most commonly adopted recommendation paradigms: DSSM~\citep{huang2013learning} and SASRec~\citep{kang2018self}.\footnote{We did not study other CTR (click-through rate) prediction  models, as they essentially belong to the same category as DSSM, with the key difference being that many CTR models are based on single-tower backbone networks~\citep{yuan2022tenrec,he2017neural,guo2017deepfm,covington2016deep}. Intuitively, such difference generally does not affect our subsequent conclusions (see section~\ref{st:common recommendation scenario}), since improvement from a two-tower backbone to a single-tower is often limited if having the same training manners~\citep{li2022inttower,zhu2022bars}.
    However, DSSM or CTR models are quite different from the seq2seq-based sequential recommendation models, such as SASRec. For example, as shown in Figure~\ref{fig:IDRec_and_MoRec}, SASRec has   $L-1$ loss functions for each interaction sequence (input: $1,2,...,L-1$, predict: $2,...,L$), while DSSM (or other CTR models)  \textit{typically}  uses one loss function to predict an interaction of a $<u, i>$ pair.}
The original DSSM model is a two-tower based architecture where users/items are encoded by their own encoder networks with user and item IDs as input. 
   SASRec is a well-known sequential recommendation model based on multi-head self-attention (MHSA)~\citep{vaswani2017attention} which describes a user by her interacted item ID sequence.  
   As mentioned before, by replacing ID embeddings
   with an item ME, 
   we obtain their MoRec versions for both DSSM and SASRec.
   We illustrate  IDRec and MoRec in Figure~\ref{fig:IDRec_and_MoRec}.

\subsection{Training Details}\label{apx:sec:recommender system frameworks}
   Denote $\mathcal{R}$ as the set of all observed interactions in the training set. For each positive $<u, i> \in \mathcal{R}$, we randomly draw a negative sample $<u, j> \notin \mathcal{R}$ in each training epoch, following~\citep{rendle2012bpr,he2017neural}. 
   The positive and sampled negative interactions can form the training set $\mathcal{R}^{train}$. Following ~\citep{he2017neural,kang2018self}, we adopt the widely used binary cross entropy loss as the objective function for both DSSM and SASRec, and their MoRec versions for a fair comparison:
    \begin{equation} \label{overallLossF}
    \left\{
    \begin{aligned}
        min
        - \sum\limits_{u \in \mathcal{U}} \sum\limits_{ i \in [2,...,L]} 
		\left\{
		\log(\sigma(\hat{y}_{ui}))
		+
		\log(1- \sigma(\hat{y}_{uj}))\right\} & \text{\,\,SASRec }\\
        min
		- \sum\limits_{<u, i, j> \in \mathcal{R}}
		\left\{
		\log(\sigma(\hat{y}_{ui}))
		+
		\log(1- \sigma(\hat{y}_{uj}))\right\} & \text{\,\,\,DSSM}
    \end{aligned}\right.
\end{equation}
 where $\sigma(x) = 1/(1+e^{-x})$ is the sigmoid function, $L$ is the interaction sequence length of user $u$. $i$ and $j$ denotes positive and negative item respectively for $u$, $\hat{y}_{ui}$ is
 the matching score between hidden vectors of user ($u$) encoder  and item ($i$) encoder. Note that SASRec's user encoder (by seq2seq training) produces a different hidden vector at each position of the interaction sequence.
 Without special mention, all parameters of the entire recommendation model 
are optimized during training in the following experiments.

\section{Experimental Setups}
   \subsection{Datasets}\label{datasets}
      
   We evaluate IDRec and MoRec on three  real-world datasets, namely, the MIND news clicks dataset from the Microsoft news recommendation platform
   ~\citep{wu2020mind}, the HM clothing purchase dataset from the H\&M   platform\footnote{https://www.kaggle.com/competitions/h-and-m-personalized-fashion-recommendations/overview} and the Bili\footnote{https://www.bilibili.com/} comment dataset from an online video recommendation platform.\footnote{To build this dataset, we randomly crawled URLs of short videos  (with duration time  less than 10 minutes)  from 23 different video channels of Bili from October 2021 to March 2022. Then we
    recorded  public comments of these videos as interactions. Finally, we  merged all user interactions chronologically and removed duplicate interactions. } 
 Purchases and comments can be considered implicit click signals, as it is reasonable to  
assume that the user has clicked on the item before making a purchase or leaving a comment.  However, we cannot assume the opposite holds, which is a common property in most recommendation datasets, i.e.  unobserved items can be either positive or negative for the user.

    \begin{figure*}[htbp] 
		\centering
              \subfigure[Item cases on ImageNet1K.] {
                         \label{apx:fig:Item cases of ImageNet1K}
                         \includegraphics[height=1.5in]{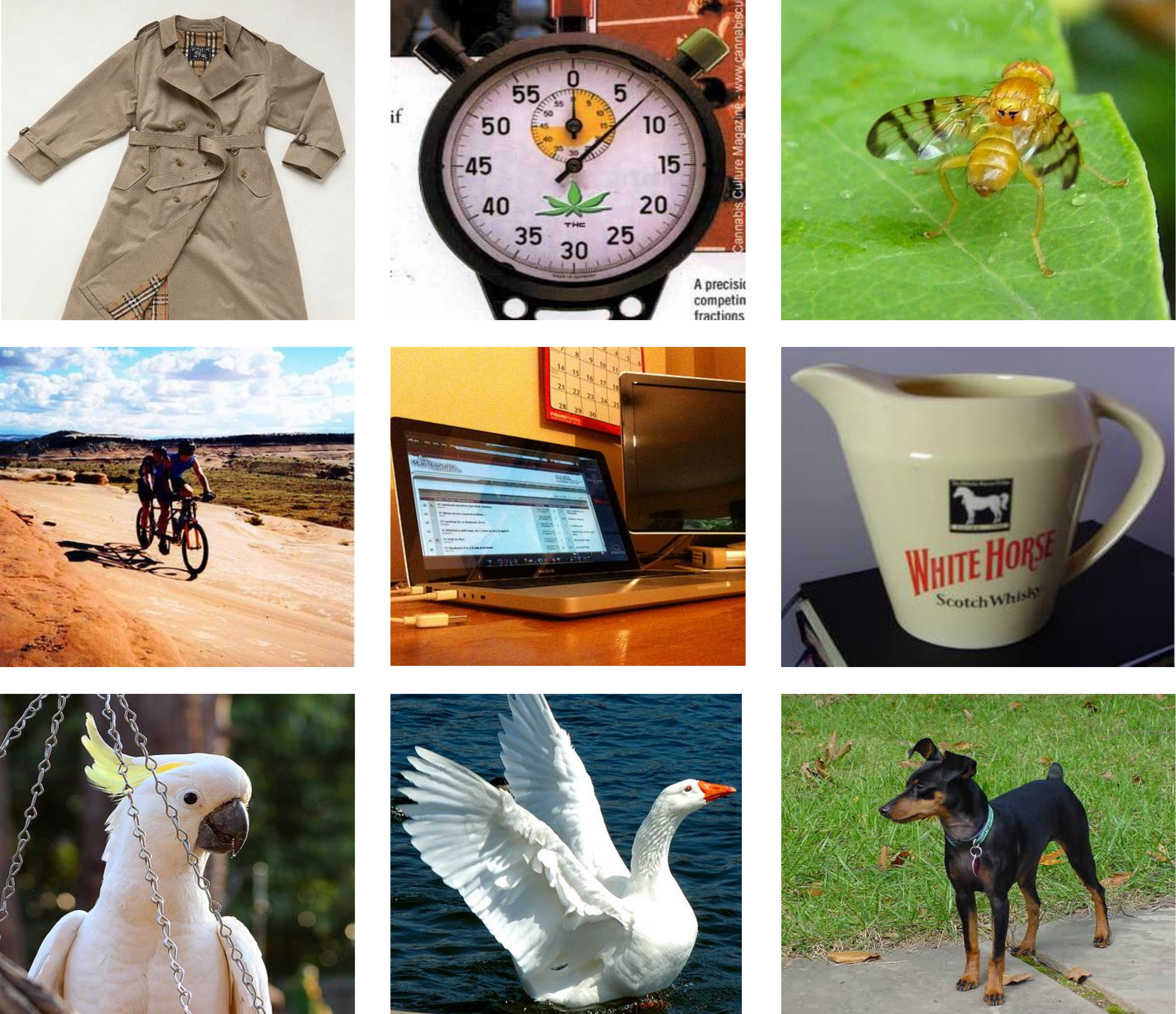} 
                          }\hspace{2mm}
              \subfigure[Item cases on HM.
              ] {
                  \label{apx:fig:Item cases of HM}
                  \includegraphics[height=1.5in]{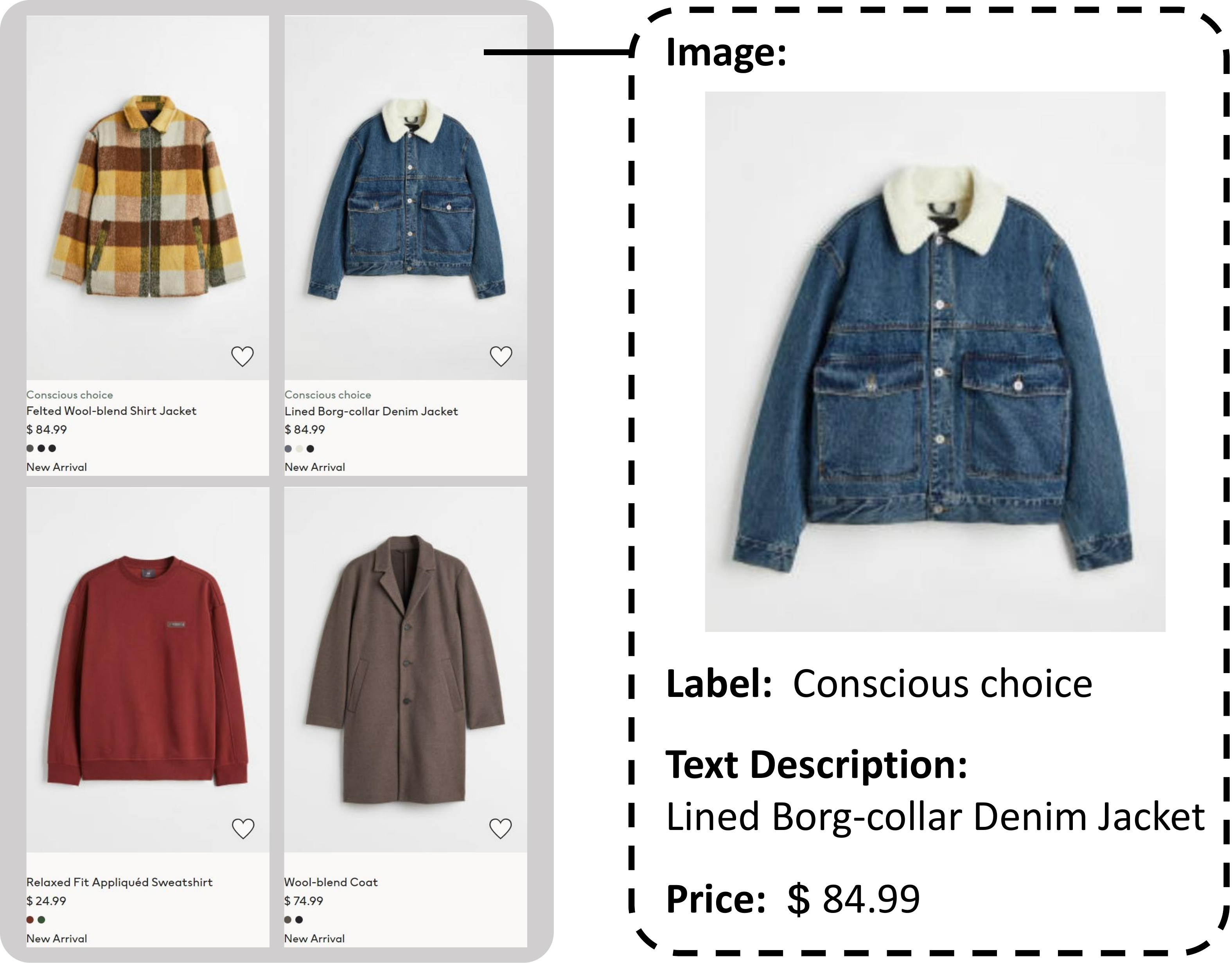}
              }\hspace{2mm}
              \subfigure[Item cases on Bili. 
              ] {
                  \label{apx:fig:Item cases of Bili}
                  \includegraphics[height=1.5in]{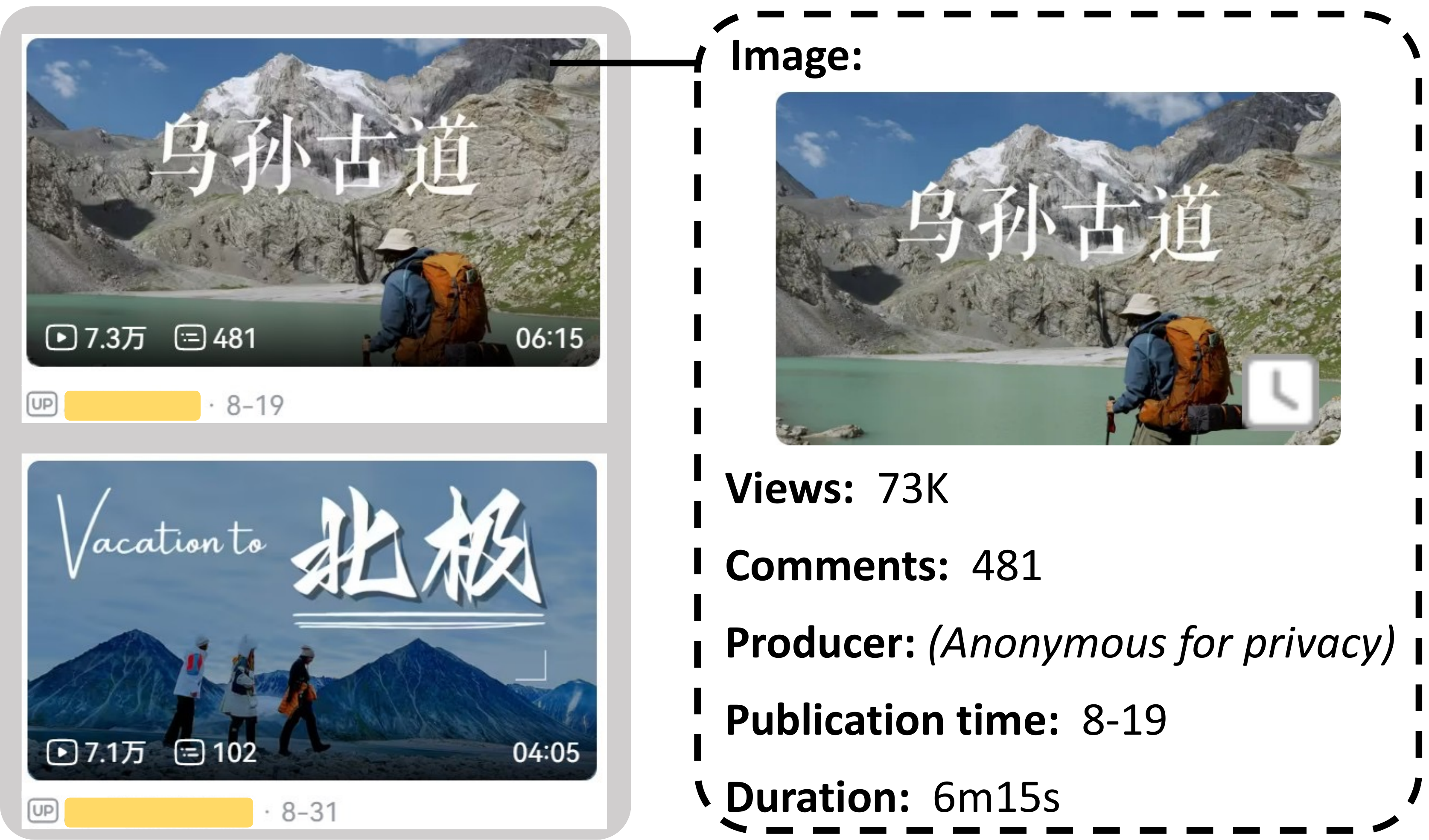}
              }
		\caption{Item cases on datasets. Image ME we used are all pre-trained in the ImageNet1K dataset.
		} \label{apx:fig:Item cases}
	\end{figure*}
 
 

To ensure a fair comparison between IDRec and MoRec,
 the dataset used should guarantee  that users' clicking decisions on an item are solely based  on the modality content features of the item. 
 Intuitively, the cover of an image or video and the title of a news article, play a crucial role in providing users with the very first impression of an item. This impression significantly influences their decision to click on the item.
 Therefore, in MIND, we represent items by their news article titles, while in HM \& Bili, we represent items using their corresponding cover images. 
  Nevertheless, it is still possible that these datasets may not \textit{perfectly}
  meet the  requirement.  Particularly, within the e-commerce context of the HM dataset,  factors such as the item's cover image, price, and sales volume may collectively influence a user's decision to click on an item (refer to Figure~\ref{apx:fig:Item cases}).
  This means relying solely on a cover image in the HM dataset may not be adequate for MoRec to effectively capture these non-visual features, as it is the only input to the item encoder.  In contrast, IDRec is known to be able to  implicitly learn such features from the latent embedding space~\citep{koren2009matrix}. That is,  MoRec's performance  may still have room for improvement if a more ideal\footnote{It seems that so far there is no publicly available dataset that fully satisfies the above mentioned requirement.}   dataset or more useful content features were taken into account.

     \begin{table}[t]
         \caption{Dataset characteristics. 
         $n$ and $m$ denote the numbers of users and items.  $|\mathcal{R}|^{train}$,  $|\mathcal{R}^{valid}|$ and $|\mathcal{R}^{test}|$ denote the number of interactions of the training set,  validation set and testing set, respectively.
         $|\mathcal{R}|/(nm)$ represents density.
         }
         \label{tb: Statistic of the datasets}
         \begin{center}
         \scalebox{0.85}{
         \begin{tabular} {c|cccccc}
         \toprule
               {Dataset} & {$n$} & {$m$} & {$|\mathcal{R}|^{train}$} & { $|\mathcal{R}^{valid}|$} & { $|\mathcal{R}^{test}|$}  &  { $|\mathcal{R}^{train}|/(nm)$ } \\ 
        \midrule
               MIND & 630$K$ & 80$K$ & 8,407$K$ & 630$K$ & 630$K$ & 0.0167\% \\
               HM & 500$K$ & 87$K$ & 5,500$K$ & 500$K$ & 500$K$  & 0.0127\% \\
               Bili & 400$K$ & 128$K$ & 4,400$K$ & 400$K$ & 400$K$ & 0.0086\% \\
        \bottomrule
                \end{tabular}}
         \end{center}
      \end{table}
      
To construct the datasets for experiments, we randomly select around  400$K$, 500$K$ and 600$K$ users from Bili, HM, and MIND, respectively. Then, we perform basic pre-processing by setting the size of all images to $224 \times224$ and the title of all news articles to a maximum of 30 tokens (covering 99\% of descriptions).
For MIND, we select the latest 23 items for each user to construct the interaction sequence. For HM and Bili, we choose the 13 most recent interactions since encoding images requires much larger GPU memory (especially with the SASRec architecture).  Following ~\citep{rendle2012bpr}, we remove users with less than 5 interactions, simply because we do not consider cold user settings in this paper.  



\subsection{Hyper-parameters}\label{apx:sec:hyper parameters and training cost}
     For all methods, we employ an AdamW~\citep{loshchilov2017decoupled} as the default optimizer and find that the dropout rate set to $0.1$ (i.e., removing 10\% parameters)  offers the optimal results on the validation set. Regarding other hyper-parameters, we follow the common practice and perform extensive  searching.
     For IDRec, \textcolor{black}{we tune the learning rate $\gamma$ from $\left\{1e\mbox{-}3, 5e\mbox{-}4, 1e\mbox{-}4, 5e\mbox{-}5\right\}$, the embedding/hidden size $d$ from $\left\{64, 128, 256, 512, 1024, 2048, 4096\right\}$. We set batch size $b$ to $1024$ for DSSM and $128$ for SASRec.} For MoRec, we set $d$ to $512$  for both DSSM and SASRec, $b$ to $512$ and $64$ for DSSM and SASRec respectively due to GPU memory constraints. Given that ME (e.g., BERT and ResNet) has already well pre-trained parameters, we use relatively smaller $\gamma$ than other parts in the recommender model. That is, we search $\gamma$  from $\left\{1e\mbox{-}4, 5e\mbox{-}5, 1e\mbox{-}5\right\}$ for the pre-trained ME networks, \textcolor{black}{and set $\gamma$ to $1e\mbox{-}4$ for other parts with randomly initialized parameters.} Finally, we tune the weight decay $\beta$ from $\left\{0.1, 0.01, 0\right\}$ for both IDRec and MoRec.
        
   For the MLPs (multilayer perceptron) used in DSSM, we initially set their middle layer size  to $d$ as well and search the layer number $l$ from $\left\{0, 1, 3, 5\right\}$ but find that $l = 0$ (i.e., no hidden layers) always produces the best results. For the Transformer block used in SASRec, we set $l$ to $2$ and the head number of the multi-head attention to $2$ for the optimal results. All other hyper-parameters are kept the same for IDRec and MoRec unless specified otherwise.
      \subsection{Comparison Settings}\label{comparisonsettings}
   For a fair comparison, we ensure that IDRec and MoRec have exactly the same network architecture except for the item encoder. For both text and vision encoders, we pass their output item representations to a DT-layer (see Figure~\ref{fig:IDRec_and_MoRec}) for dimension transformation. Regarding the hyper-parameter setting, our principle is to ensure that IDRec are fully tuned in terms of learning rate $\gamma$, embedding size $d$, layer number $l$, and dropout $\rho$. While for MoRec, we attempt to first use the same set of hyper-parameters as IDRec and then perform some basic searching around the best choices. Therefore, without special mention, we do not guarantee that the results reported by MoRec are the best, because searching all possible hyperparameters for MoRec is very expensive and time-consuming,  sometimes taking more than 100x compute and training time than IDRec, especially for vision, see Table~\ref{tb: training cost}.  Thereby, how to efficiently find the optimal hyper-parameters of MoRec is an important but unexplored research topic.

\begin{figure*}[t] 
               \vspace{-0.1in}
		\begin{minipage}[htbp]{0.77\linewidth}
			\centering
			\tabcaption{Accuracy (\%)  comparison of IDRec and MoRec using DSSM and SASRec for regular setting.   MoRec with different ME are directly denoted by their encoder names for clarity. 
   The best results for DSSM and SASRec are bolded.
   `Improv.' is the relative improvement of the best MoRec compared with the best IDRec.
   All results of MoRec are obtained by fine-tuning their whole parameters  including both the item encoder and user encoder. Swin-T and Swin-B are Swin Transformer with different model sizes, where T is tiny and B is base. ResNet50 is a 50-layer ResNet variant.}
        	\label{tb:common item recommendation}
			\scalebox{0.85}{
               \begin{tabular}{cc|cccccccc}
               \toprule
               \multirow{2}{*}{Dataset} & \multirow{2}{*}{Metrics} &\multicolumn{3}{c}{DSSM}  
               &\multicolumn{4}{c}{SASRec} & \multirow{2}{*}{Improv.}\\
               \cmidrule(lr){3-5}\cmidrule(lr){6-9}
               & 
               &IDRec &$\text{BERT}_{\text{base}}$ &$\text{RoBERTa}_{\text{base}}$
               &IDRec &$\text{BERT}_{\text{small}}$ &$\text{BERT}_{\text{base}}$ &$\text{RoBERTa}_{\text{base}}$ &\\ 
               \midrule
               \multirow{2}*{MIND}
               &HR@10
               &\textbf{3.58} &2.68 &3.07
               &17.71  &18.50 &18.23  &\textbf{18.68}   &+5.48\%  \\
               &NDCG@10 
               &\textbf{1.69} &1.21  &1.35
               &9.52  &9.94 &9.73 &\textbf{10.02} &+5.25\%  \\
               \midrule
               &
               &IDRec &ResNet50 &Swin-T
               &IDRec &ResNet50 &Swin-T &Swin-B & \\ 
               \midrule
               \multirow{2}*{HM}
               &HR@10
               &\textbf{4.93}  &1.49 &1.87
               &6.84	  &6.67 &6.97 &\textbf{7.24}  &+5.85\% \\  %
               &NDCG@10 
               &\textbf{2.93}  &0.75 &0.94
               &\textbf{4.01}  &3.56 &3.80  &3.98 &-0.75\% \\   %
               \midrule
               \multirow{2}*{Bili}
               &HR@10
               &\textbf{1.14} & 0.38  &0.57   
               &3.03	 & 2.93  &3.18  &\textbf{3.28}  &+8.25\%\\
               &NDCG@10 
               &\textbf{0.56}  &0.18  &0.27  
               &1.63 &1.45  &1.59  &\textbf{1.66}  &+1.84\%\\
               \bottomrule
               \end{tabular}}
		\end{minipage}
   \hspace{0.1in}
		\begin{minipage}[htbp]{0.2\linewidth}
			\centering
               \vspace{0.1in}
                \includegraphics[width=1.4in]{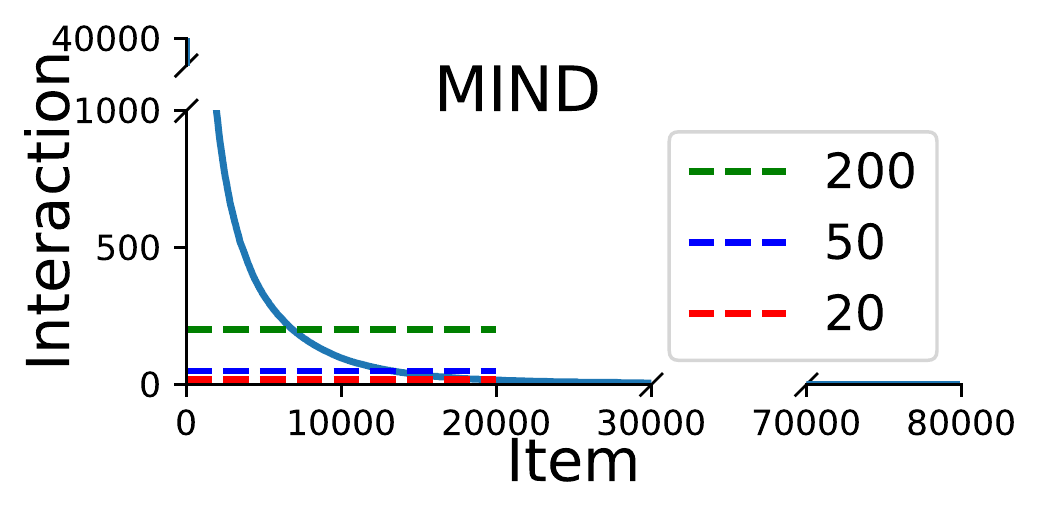}
               \includegraphics[width=1.4in]{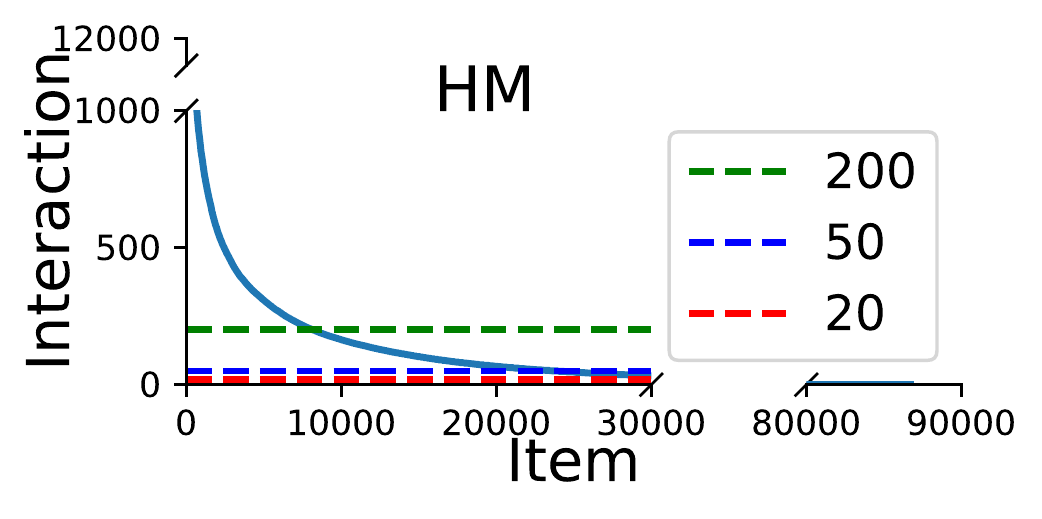}
               \includegraphics[width=1.4in]{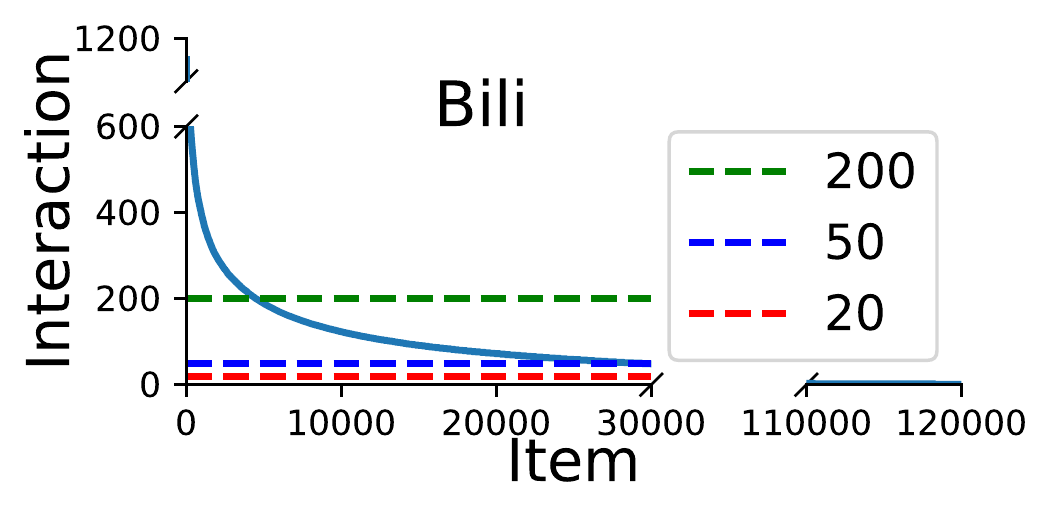}
	        \figcaption{Item popularity distribution.}
	        \label{fig:mind_long_tail.pdf}
            \label{fg:Item popularity distribution}
		\end{minipage}
\end{figure*}

    \subsection{Evaluations}
    We split the datasets into training, validation, and testing sets by adopting the standard leave-one-out strategy. Specifically, the latest interaction of each user was used for evaluation, while second-to-last was used as validation for hyper-parameter searching, and all others are used for training.
   We evaluate all models using two popular top-N ranking metrics: HR@N (Hit Ratio) and NDCG@N (Normalized Discounted Cumulative Gain), where N is set to $10$. 
   We rank the ground-truth target item by comparing it with all the left items in the item pool. Finally,
   we report results on the testing set, but find the best hyper-parameters via the validation set.

\section{Comparative Studies (Q(i))}
According to existing literature, MoRec can easily beat IDRec in the new item or cold-start item recommendation settings~\citep{tang2019adversarial,packer2018visually,hansen2020content}. We report such results in the Appendix~\ref{morecvsidrec}.
In this paper we focus  on evaluating them in the more challenging setting: regular (mixture of warm and cold items) and warm-start item recommendation scenarios, where IDRec is usually very strong. To the best of our knowledge, such comparisons have not been explicitly discussed in the existing literature. 

 As mentioned, we evaluate IDRec and MoRec with the two most important recommendation architectures, i.e.,
   DSSM and SASRec. We use  pre-trained BERT and RoBERTa as ME when items are of text features, and use pre-trained ResNet and Swin Transformer~\citep{liu2021swin}  when items are of visual features.\footnote{We provide the  URLs of all pre-trained modality encoders utilized in our study at  \textcolor{blue}{https://github.com/westlake-repl/IDvs.MoRec}.}
   Note for BERT and RoBERTa, we add the DT-layer (see Figure~\ref{fig:IDRec_and_MoRec}) on the final representation of the `[CLS]' token.
   We report results on the testing set in Table~\ref{tb:common item recommendation} for regular setting (i.e. the original distribution) and Table~\ref{tb:warmrec} for warm-start settings where cold items are removed. 

    \subsection{MoRec \emph{vs} IDRec (Regular Setting)}
    \label{st:common recommendation scenario}
   As shown in Table~\ref{tb:common item recommendation}, we observe that DSSM always substantially underperforms SASRec, regardless of the item encoding strategy used. For instance,
   SASRec-based IDRec is around  4.9$ \times $ better than DSSM-based IDRec in terms of HR@10 for news recommendation, although their training, validation, and testing sets are  kept exactly the same. The performance gap for image recommendation is relatively small, around 1.4$\times$ and 2.7$\times$,  on HM and Bili, respectively.
   This is consistent with much  prior literature~\citep{hidasi2015session,kang2018self}, where
    representing and modeling users with their interacted item sequence 
  is often more powerful than dealing them as individual userIDs. 
   
   Second, we notice that with the DSSM architecture, MoRec perform much worse than 
   IDRec in all three datasets even with the state-of-the-art (SOTA) ME, in particular for the visual recommendation scenarios. By contrast,  with the SASRec architecture, MoRec consistently achieve better results than IDRec on MIND using any of the three text encoders, i.e., $\text{BERT}_{\text{small}}$, $\text{BERT}_{\text{base}}$ and $\text{RoBERTa}_{\text{base}}$. For instance, MoRec outperform IDRec by over 5\% on the two evaluation metrics with the $\text{RoBERTa}_{\text{base}}$ text encoder.  Meanwhile, MoRec perform comparably to IDRec when using Swin Transformer as ME but perform relatively worse when using ResNet50.
   The performance disparity of MoRec between DSSM and SASRec potentially implies that \textbf{a powerful 
recommendation  backbone (SASRec\emph{vs} DSSM ) and training approach (seq2seq\emph{vs} $<u, i>$ pair)  is required to fully harness the strengths of the modality-based item  encoder}. Given MoRec's poor results with DSSM, we mainly focus on the SASRec architecture in the following.

\begin{table}[t]
   \caption{MoRec \emph{vs} IDRec (HR@10) in the warm-start settings with SASRec as user backbone. Warm-20 means removing items with less than 20 interactions in the original dataset.}
   \label{tb:warmrec}
   \begin{center}
   \scalebox{0.85}{
   \begin{tabular}{c|cc   cc  cc}
   \toprule
   \multirow{2}{*}{Dataset} &\multicolumn{2}{c}{MIND}
    &\multicolumn{2}{c}{HM} 
     &\multicolumn{2}{c}{Bili} \\
   \cmidrule(lr){2-3}\cmidrule(lr){4-5}\cmidrule(lr){6-7}
   &IDRec &$\text{BERT}_{\text{base}}$ 
   &IDRec &Swin-T &IDRec&Swin-T \\ 
   \midrule
   
   Warm-20  & 20.12 &\textbf{20.19} 
   & 7.89  &\textbf{8.05}   
   &3.48  &\textbf{3.57}  \\ 
   
   Warm-50  & 20.65 &\textbf{20.89}    
   &\textbf{8.88}    &8.83    
   &\textbf{4.04}  &4.02  \\ 
   
   Warm-200 & \textbf{22.00} &21.73   
   &\textbf{11.15}   &11.10       
   &\textbf{10.04} &9.98 \\ 
   \bottomrule
   \end{tabular}}
   \end{center}
   \end{table}
   
\subsection{MoRec \emph{vs} IDRec (Warm Item Settings)}
\label{st:hot-sc}
    To validate the performance of MoRec and IDRec for warm item recommendation, we constructed new datasets with different item popularity. We show the item popularity distribution of the original datasets in Figure~\ref{fg:Item popularity distribution}. For each dataset, we remove items with less than 20, 50, 200 interactions from the original datasets.
 We report the recommendation accuracy of all three datasets in Table~\ref{tb:warmrec}.
It can be seen that IDRec is getting stronger and stronger from warm-20, warm-50 to warm-200. In warm-20 dataset, MoRec is slightly better than IDRec, while in warm-200, MoRec is slightly worse than IDRec for text recommendation. This is reasonable since IDRec is known to to be good at modeling popular items according to the existing literature~\citep{chen2021autodebias,yi2019sampling,yuan2016lambdafm}. But even in these warm-start setting, MoRec is still comparable to IDRec. Such property is appealing since it is well-known that MoRec can easily beat IDRec in the cold-start setting (see Appendix) and has a natural advantage for tranfer learning or cross-domain recommendation. Even further, recent work have shown that large MoRec models have the potential to be a foundation recommendation models~\citep{shin2021scaling,shin2021one4all}, capable of achieving the ambitious goal of ``one model for all''~\citep{shin2021one4all,wang2022transrec}.
 

  The above results shed the following insights: (1) the  recommendation architecture (seq2seq SASRec or two-tower DSSM)  of MoRec has a very large impact on its performance; (2) its item ME also influences the performance of  MoRec;
   (3)  (\textbf{Answer for Q(i)}) \textbf{ equipped with the most powerful ME, MoRec can basically beat its IDRec counterpart for text recommendation (both cold and warm item settings)  and is on par with IDRec for visual recommendation  when using the sequential neural network recommendation architecture. However, it seems that
   there is little chance for MoRec to replace IDRec with the typical DSSM  training approach in either  regular or the warm-start setting;} 
   (4) although MoRec cannot beat IDRec in terms of very popular item recommendation, they still show very competitive results. 
   To the best of our knowledge, this is the first paper that explicitly claims that pure MoRec can be comparable to pure IDRec (when they are compared under the same sequential\footnote{Due to space reasons, we do not report results of other sequential models, but we have indeed evaluated them. The conclusions hold when using GRU4Rec, NextItNet, and BERT4Rec as backbones.} recommendation architecture),  even for the very challenging warm item recommendation.

   \section{
   Inherit Advances in NLP \& CV?  
   (Q(ii))
   }
   \label{do MoRec benefit from}
Intuitively, MoRec have the potential to bring powerful representation learning techniques from other communities, such as NLP and CV, to recommendation tasks. However, this has not been formally studied.
   Here, we ask: can recent advances in NLP and CV translate into improved accuracy for recommendation tasks?  We aim to answer it from the following perspectives.

\begin{figure}[t]
      \centering
      \includegraphics[width=3.35in]{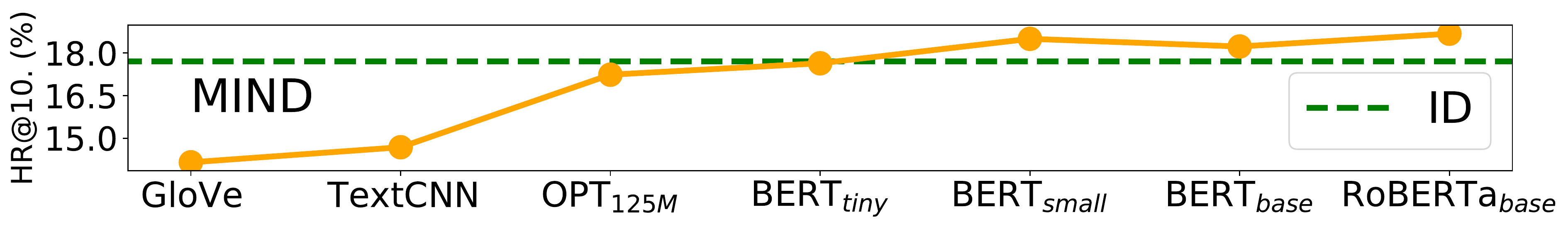}
      \includegraphics[width=3.2in]{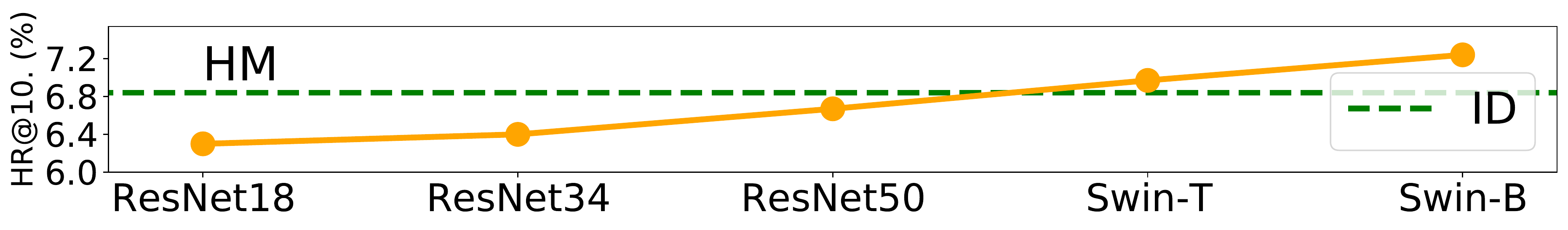}
      \includegraphics[width=3.2in]{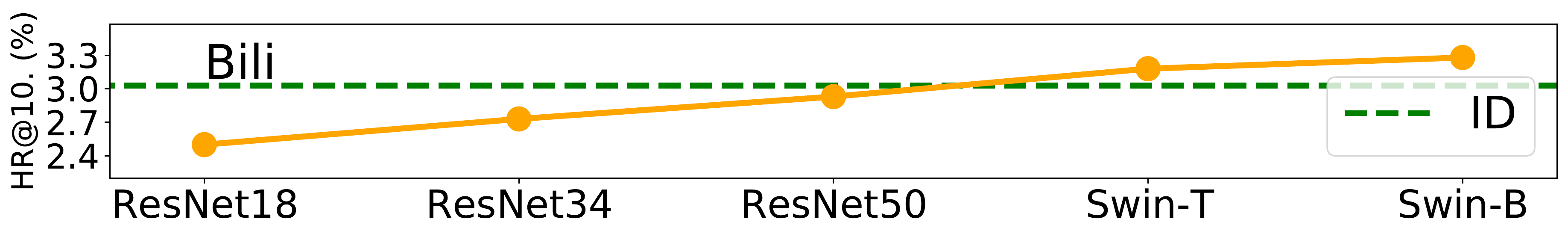}
      \caption{Accuracy with different pre-trained ME in MoRec.  Parameters of the pre-trained encoder network are all fine-tuned on the recommendation task.}
      \label{fig:lager and advanced}
    \end{figure}

  First, we investigate whether a larger pre-trained ME enables better recommendation accuracy since in NLP and CV larger pre-trained models tend to offer higher performance in corresponding downstream tasks. As shown in Figure~\ref{fig:lager and advanced}, a larger vision item encoder always achieves better image recommendation accuracy, i.e., ResNet18-based MoRec $<$ ResNet34-based MoRec  $<$ ResNet50-based MoRec, and Swin-T based MoRec $<$  Swin-B based MoRec. Similarly, we find that $\text{BERT}_{\text{tiny}}$-based MoRec $<$ $\text{BERT}_{\text{base}}$-based MoRec $<$ $\text{BERT}_{\text{small}}$-based MoRec. One difference is that $\text{BERT}_{\text{base}}$-based MoRec do not outperform $\text{BERT}_{\text{small}}$-based MoRec although the latter has a smaller-size BERT variant.  We conclude that,
  in general, a larger and more powerful ME from NLP and CV tends to improve the recommendation accuracy, but this may not strictly apply in all cases.
    
  Second, we investigate whether a stronger encoder network enables better recommendations.
  For example, it is recognized that RoBERTa  outperforms BERT~\citep{liu2019roberta}, and BERT outperforms the unidirectional GPT~\citep{radford2018improving}, such as OPT~\citep{zhang2022opt},  for most NLP understanding  (but not generative) tasks with similar model sizes, and that Swin Transformer often outperforms ResNet in many CV tasks~\citep{liu2021swin}. 
 In addition, these modern pre-trained NLP foundation models easily outperform TexTCNN~\citep{kim-2014-convolutional} and GloVe~\citep{pennington2014glove}, two well-known shallow models developed about ten years ago.
  As shown in Figure~\ref{fig:lager and advanced}, MoRec's performance keeps consistent
  with the findings in NLP and CV, i.e., $\text{RoBERTa}_{\text{base}}$-based MoRec $>$ $\text{BERT}_{\text{base}}$-based MoRec $>$ $\text{OPT}_{\text{125M}}$-based MoRec $>$ TextCNN-based MoRec $>$ GloVe-based MoRec, and Swin-T based MoRec $>$ ResNet50-based MoRec (Swin-T has a similar model size to ResNet50, the same for $\text{RoBERTa}_{\text{base}}$, $\text{BERT}_{\text{base}}$ and $\text{OPT}_{\text{125M}}$).

  Third, we investigate whether the pre-trained ME produces higher recommendation accuracy than its training-from-scratch (TFS) version (i.e., with random initialization). There is no doubt that the pre-trained BERT, ResNet, and Swin largely improve corresponding NLP and CV
  tasks against their TFS versions.
  We report the recommendation results on the testing set in Table~\ref{tab:training form scratch}. It can be clearly seen that pre-trained MoRec obtain better final results. In particular, MoRec achieve around 10\% improvements with the pre-trained ME (ResNet and Swin) on HM and Bili, which also aligns with findings in NLP and CV domains.
  We also construct the smaller version datasets by randomly drawing 50K users from MIND, HM, and Bili. It can be seen that the advantages of pre-trained ME over TFS are more obvious on small datasets. However, we found that the pre-trained $\text{BERT}_{\text{base}}$ is even worse than its TFS version on MIND-50K.

\begin{table}[t]
         \caption{Pre-trained (PE) ME \emph{vs} TFS on the testing set regarding HR@10 (\%). $\text{BERT}_{\text{base}}$ are used as text ME, and ResNet50 and Swin-T are used as vision ME. `Improv.' indicates the relative improvement of PE over TFS.
         }
         \label{tab:training form scratch}
         \begin{center}
         \scalebox{0.85}{
                \begin{tabular}{ c c| c c c c c c }
               \toprule
               \multirow{2}{*}{Dataset} &\multirow{2}{*}{ME}  &\multicolumn{3}{c}{Base}&\multicolumn{3}{c}{50K}\\
               \cmidrule(lr){3-5}\cmidrule(lr){6-8}
                & &TFS   &PE  &Improv.  &TFS   &PE  &Improv.\\
               \midrule
                MIND &$\text{BERT}_{\text{base}}$ &17.78  &\textbf{18.23} &+2.53\% &\textbf{15.04}&14.35 & -4.59\%\\
               \midrule
                \multirow{2}{*}{HM}  &ResNet50    &5.82 &\textbf{6.67} &+14.60\% &2.74 &\textbf{3.26} &+18.98\%\\
                  &Swin-T    &6.27 &\textbf{6.97} &+11.16\% &2.84 &\textbf{4.47} &+57.39\% \\
               \midrule
                \multirow{2}{*}{Bili} &ResNet50  &2.67  &\textbf{2.93} &+9.74\%  &1.07  &\textbf{1.20} &+12.05\% \\
                 &Swin-T    &2.83  &\textbf{3.18} &+12.37\% &1.08  &\textbf{1.46} &+35.19\% \\
               \bottomrule
               \end{tabular}}
         \end{center}
      \end{table}

  According to the above experiments, we conclude that \textbf{ (Answer for Q(ii)) MoRec build connections for RS and other multimedia  communities, and can in general inherit the latest advances from the NLP and CV fields}. This is a very good property, which means  that   once there are new breakthroughs in the corresponding research fields in the future, MoRec have more opportunities and greater room to be  improved.
  
   \section{Are modality representations universal for RS?
   (Q(iii))}\label{sec:freezing vs fine-tuning}
   Foundation models in NLP and CV are expected to generate generic representation, which can then be directly used for downstream tasks in the zero-shot setting. However, most of them are only evaluated in some traditional tasks~\citep{radford2021learning,li2022grounded}, such as image  and text classification. We argue that predicting user preference is more challenging than these objective tasks.

To see this problem clearly, we evaluate two training approaches.
The first approach is to pre-extract modality features by ME and then add them into a recommendation model ~\citep{he2016vbpr,he2016ups}, referred to as a two-stage (TS) pipeline. Due to the high training efficiency, TS is especially popular in real-world industrial applications, where there are usually hundreds of millions of training examples.
The second approach is the one used in all above experiments, by optimizing user and item encoders simultaneously in an E2E manner.
   

As shown in Table~\ref{tab:2stage_end2end}, we find that TS-based MoRec show surprisingly poor results, compared to IDRec and E2E-based MoRec. In particular, with ResNet, it achieves only around 60\% and 25\% performance of E2E-based MoRec on HM and Bili, respectively. For better adaption, we also add many dense layers on top of these fixed modality features. As shown, this can indeed improve the performance of TS; however, it is still much worse than IDRec and E2E-based MoRec, especially for visual recommendation.
 
     The results indicate that the modality features learned by these NLP and CV tasks are not universal enough for the recommendation problem, and thus the recommendation results are worse compared to retraining on new data (i.e., the E2E paradigm).  The good thing is that by proper adaption (i.e., TS-DNN), TS-based MoRec have some potential to compete  with E2E MoRec for text recommendation in the future (16.66 \emph{vs} 18.23).
     



    Thereby, we want to explicitly remind RS researchers and practitioners that \textbf{(Answer for Q(iii)) the popular two-stage recommendation mechanism leads to significant performance degradation (especially for image recommendation), which should not be ignored in practice.\footnote{Unfortunately, so far, there is not even any literature showing that an E2E-based MoRec has been successfully deployed in real-world recommender systems. 
    } Second, for NLP and CV researchers, we want to show them that, despite the revolutionary success of FM, until now their representation features are not universal enough, at least for item recommendation.}

   \section{key challenges (Q(iv))}
   E2E-based MoRec has been less studied before, especially for visual recommendation. Here, we present several key challenges and some unexpected findings that the community may not be aware of.

   \begin{table}[t] 
         \caption{HR@10 (\%) of E2E \emph{vs} TS with additional MLP layers . `TS-DNN 6' denotes that TS-based MoRec with 6 learnable MLPs layers on top of these fixed modality representation.
         }
         \label{tab:2stage_end2end}
         \begin{center}
         \scalebox{0.8}{
         \begin{tabular}{ p{0.8cm}<{\centering} p{0.6cm}<{\centering} p{1.1cm}<{\centering}| c c c c c c   c}
               \toprule
               \multirow{2}{*}{Dataset} &\multirow{2}{*}{IDRec}  &\multirow{2}{*}{ME} &\multirow{2}{*}{TS}  &\multicolumn{5}{c}{TS-DNN} &\multirow{2}{*}{E2E}  \\
               \cmidrule(lr){5-9}
               & & &  &2  & 6 &8 &10 &12 & \\
               \midrule
                MIND &17.71 &$\text{BERT}_{\text{base}}$
                &13.93 &15.20 &16.26  &\underline{16.66}  &16.32 &16.14  &\textbf{18.23} \\
               \midrule
                \multirow{2}{*}{HM} &\multirow{2}{*}{6.84}
                &ResNet50  &4.03 &4.64 &\underline{5.40}  &5.39 &\underline{5.40} &5.02 &6.67\\
                &&Swin-T   &3.45 &4.46 &5.28 &\underline{5.55} &5.40 &5.38 &\textbf{6.97} \\
               \midrule
                \multirow{2}{*}{Bili} &\multirow{2}{*}{3.03}
                &ResNet50  &0.72 &1.23 &\underline{1.62} &1.47  &1.28&1.24 &2.93\\
                &&Swin-T    &0.79 &1.40 &1.81 &\underline{2.10}  &1.95&1.64 &\textbf{3.18} \\
               \bottomrule
               \end{tabular}}
         \end{center}
      \end{table}

  \begin{table}[t]
   \caption{The training cost. \#Param: number of tunable parameters, FLOPs: computational complexity (we measure FLOPs with batchsize=1), Time/E: averaged training time for one epoch, `m' means minutes, MU: GPU memory usage, e.g., `V100-32G(2)' means that we used 2 V100s with 32G memory.  
  }
   \label{tb: training cost}
   \begin{center}
   \scalebox{0.83}{
  \begin{tabular}{ c l | ccccc}
   \toprule
    Dataset & Method 
  &\#Param. &FLOPs &Time/E &MU &GPU \\ 
   \midrule
    \multirow{4}*{MIND}   &IDRec     &47M  &0.12G &2.7m     &3G   &V100-32G(1)\\
    &$\text{BERT}_{\text{tiny}}$    &11M  &0.63G &10m  &4G   &V100-32G(1)\\
    &$\text{BERT}_{\text{small}}$   &35M  &16G   &42m    &13G  &V100-32G(1)\\
    &$\text{BERT}_{\text{base}}$     &116M &107G &102m   &52G  &V100-32G\textbf{(2)}\\
   \midrule
    \multirow{6}*{HM}     &IDRec   &114M &1G &4.3m     &5G   &V100-32G(1)\\
    &ResNet18                      &18M  &40G   &95m     &23G  &V100-32G(1)\\
    &ResNet34                      &29M  &81G   &136m    &30G  &V100-32G(1)\\
    &ResNet50                       &31M  &91G   &83m    &80G  &V100-32G\textbf{(4)} \\
    &Swin-T                         &34M  &96G   &107m    &157G &A100-40G\textbf{(4)} \\
    &Swin-B                         &94M  &333G  &102m    &308G &A100-40G\textbf{(8)}\\
   \bottomrule
   \end{tabular}}
   \end{center}
   \end{table}

\begin{table*}[t]  
   \caption{HR@10 (\%) of co-training ID and modality. `ADD' and `CON' are two fusion methods. w/ and w/o denote whether to add extra MLP layers after the fusion layer. We search the layer number from $\left\{2, 4, 6, 8\right\}$.
   Adding extra DNN layers for `ID+E2E' does not improve the accuracy, so we do not report them below for clarity.
   `Improv.' means the relative improvement with ID+modality  featurs compared to the best result of pure IDRec and pure MoRec.} 
   \label{tb:ID+Mo}
   \begin{center}
   \scalebox{0.82}{
   \begin{tabular}{cc|c|cccccc|cccccc|cccc}
   \toprule
   \multirow{3}{*}{Dataset} &\multirow{3}{*}{ME}&\multirow{3}{*}{IDRec}  &\multirow{3}{*}{TS}
   &\multicolumn{4}{c}{ID+TS}&\multirow{3}{*}{Improv.} &\multirow{3}{*}{TS-DNN} &\multicolumn{4}{c}{ID+TS-DNN}&\multirow{3}{*}{Improv.}
   &\multirow{3}{*}{E2E} &\multicolumn{2}{c}{ID+E2E}&\multirow{3}{*}{Improv.}\\
   \cmidrule(lr){5-8}\cmidrule(lr){11-14}\cmidrule(lr){17-18}
   &&&&\multicolumn{2}{c}{w/o}&\multicolumn{2}{c}{w/}&&&\multicolumn{2}{c}{w/o}&\multicolumn{2}{c}{w/}&&&\multicolumn{2}{c}{w/o}\\
    & & & &ADD & CON &ADD & CON  & & &ADD & CON &ADD & CON  && &ADD & CON   \\
   \midrule
   MIND &$\text{BERT}_{\text{base}}$ &17.71 &13.93 &16.10 &17.20 &\underline{17.66}  &17.57  &-0.28\%  &16.66 &14.93&16.58  &17.29 &\underline{17.55}  &-0.90\% &\textbf{18.23} &16.25 &\underline{17.12} &-6.09\%\\
   HM  &Swin-T &6.84  &3.45   &\underline{5.75}  &4.89  &5.37 &5.40 &-15.94\% &5.55 &\underline{5.27} &4.00 &4.77&5.11 &-22.95\%  &\textbf{6.97} &\underline{5.40} &4.95 &-22.53\%    \\
   Bili&Swin-T &3.03  &0.79 &3.01  &2.61  &\underline{3.02} &2.86 &-0.33\% &2.10  &\underline{2.86} &2.35 &2.50&2.72&-5.61\%&\textbf{3.18}  &\underline{2.94} & 2.55   &-7.55\% \\
   \bottomrule
   \end{tabular}}
   \end{center}
   \end{table*}

    \textbf{Training cost.}
As shown in Figure~\ref{fig:lager and advanced}, MoRec with larger ME tend to perform better than smaller ME, however, the training compute, time and GPU memory consumption also increase, especially for the seq2seq-based architecture with very long interaction sequence.
    We report the training cost details on HM (close to Bili) and MIND
    in Table~\ref{tb: training cost}. In fact,
    it is not difficult to imagine that MoRec will consume more computing resources and time than IDRec. However, it is hard to imagine that the  best MoRec (with SASRec as user encoder and Swin-B as ME) takes an astonishing more than 100x compute and training time than IDRec.\footnote{Note that the inference time of MoRec for online service is as fast as IDRec.}
    This has not been explicitly revealed in literature. 
    This may also be the reason why there are no formal publications combining seq2seq user encoder and E2E-learned item ME for MoRec, especially for image recommendation. Note that in practice, it may not always be necessary to optimize all parameters of ME, and for some datasets, fine-tuning a few top layers of ME can achieve comparable results.
    On the other hand, although E2E-based MoRec is highly expensive\footnote{This entire work has costed us over \$140,000. For example,  with 8 A100 GPUs, MoRec with Swin-B requires nearly 1 week to converge on HM, and
   the cost of purchasing the GPU service is about \$2,000 (for one set of hyper-parameters). } during training  (akin to FM in NLP and CV), it has been shown to enable foundation recommendation models, which can free up more labor in training specific models~\citep{shin2021one4all,shin2021scaling}. 

   
   \textbf{Extra pre-training.}
   Performing a second round of pre-training for ME using the downstream dataset  often works well in much machine learning literature~\citep{sun2019fine,gururangan2020don}. Here, we explore whether it offers improved results for MoRec.
   Following the pre-training of BERT, we adopt the “masked language model” (MLM)  objective to train the text encoder of MoRec (denoted by $\text{BERT}_{\text{base}}$-MLM) on MIND and report results in Table~\ref{tb:sendstagepretraining}. 
   As shown, $\text{BERT}_{\text{base}}$-MLM gains higher accuracy than $\text{BERT}_{\text{base}}$ for both the TS and E2E models. 
   Similarly, we explore whether it holds for the vision encoder. Note that ResNet and Swin Transformer used in previous experiments are pre-trained in a supervised manner, but neither HM nor Bili contains supervised image labels. To this end, we turn to use MAE~\citep{he2022masked}, a SOTA image encoder pre-trained in an unsupervised manner, similar to MLM.  We find $\text{MAE}_{\text{base}}$-MLM clearly improves the standard $\text{MAE}_{\text{base}}$ on HM with the TS model, but  obtains marginal gains with the E2E model. By contrast, no accuracy improvements are observed on Bili. By examining
  image cases in Figure~\ref{apx:fig:Item cases}, we find that pictures in Bili have very diverse topics and are more challenging than HM (with only very simple fashion elements).
   Our conclusion is that the effectiveness of the second round of pre-training depends on individual datasets; more importantly, it seems difficult to achieve larger accuracy gains for the E2E MoRec.

\textbf{Combing ID \& modality features.}
Given that  IDRec and E2E-based MoRec both work well, a natural idea is to combine the two features (i.e., ID and modality) in one model. We have evaluated this, as shown in Table~\ref{tb:ID+Mo}. We consider two types of feature combinations: additive and concatenated. Surprisingly, we find that neither TS- nor E2E-based MoRec is improved  compared to the best results between IDRec and MoRec. By adding ID features, E2E-based MoRec  perform even worse than pure IDRec and pure MoRec.
Our results here are somewhat inconsistent with previous publications, which often claimed to achieve better results by adding modality or multimedia features for IDRec~\citep{kula2015metadata,he2016vbpr,he2016ups}.
One reason might be that in the regular (\emph{vs} cold-start) setting, both E2E-based MoRec and IDRec learn user preference from user-item interaction data, so they cannot complement each other, while for TS-based MoRec, since ID embeddings are too much better than frozen modality features, their combination also does not improve the results.
The second reason may be that more advanced techniques are required when combining ID and modality features. %
In fact, from another point of view, MoRec with ID features will lose many advantages of MoRec (see Introduction). For example, with ID features MoRec are not suitable for building foundation recommendation models, because IDs are not easily transferable due to privacy and overlapping issues.\footnote{In this paper, we did not intend to study the effect of transfer learning, because a reliable pre-trained  model requires a huge amount of training data and compute, see~\citep{shin2021scaling,shin2021one4all}.}  %

\begin{figure}[t] 
       \subfigure[MoRec with $\text{BERT}_{\text{base}}$ on MIND.] {
           \label{fig:learning_rate_mind}
           \includegraphics[width=1.5in]{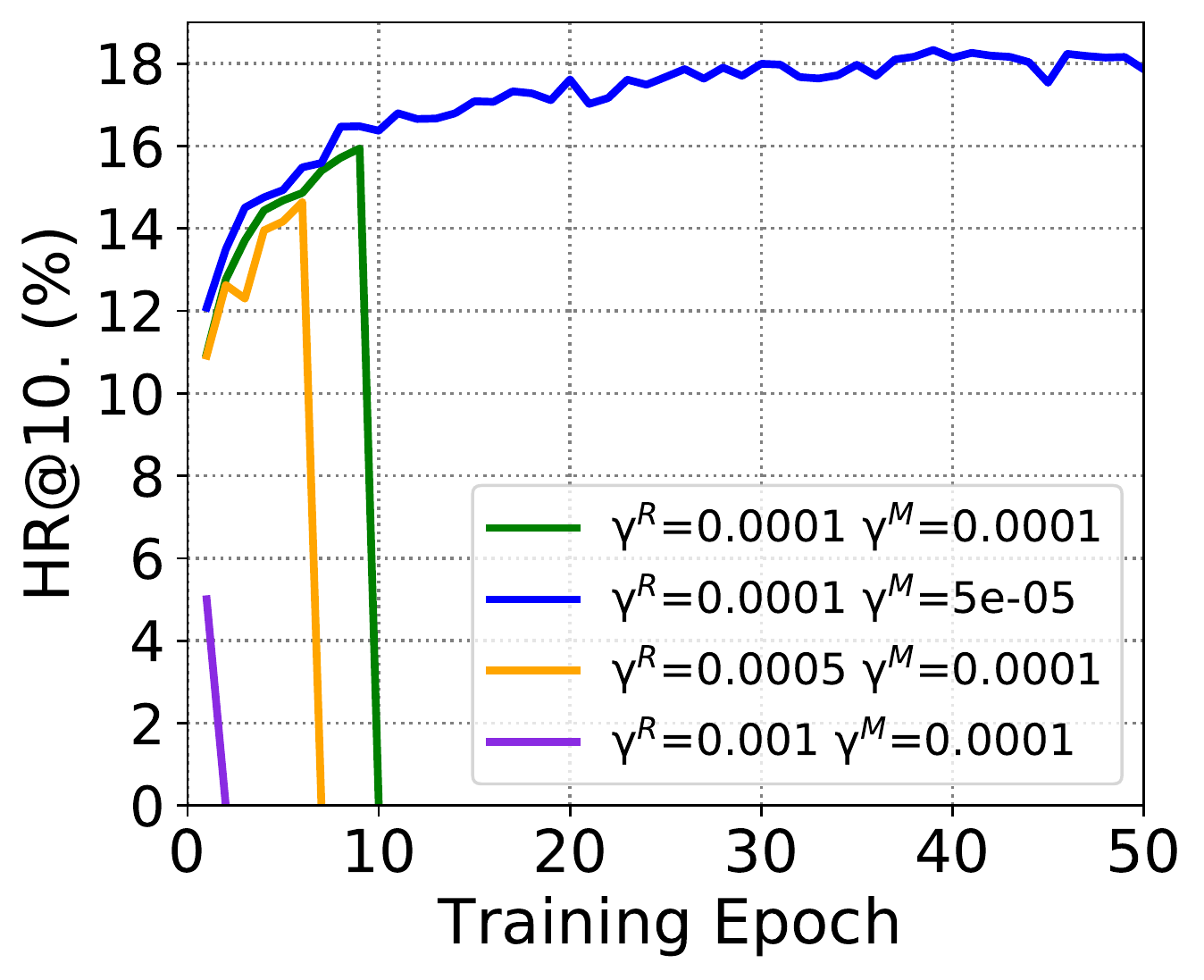}
       }
       \subfigure[MoRec with Swin-T on HM.] {
           \label{fig:learning_rate_hm}
           \includegraphics[width=1.5in]{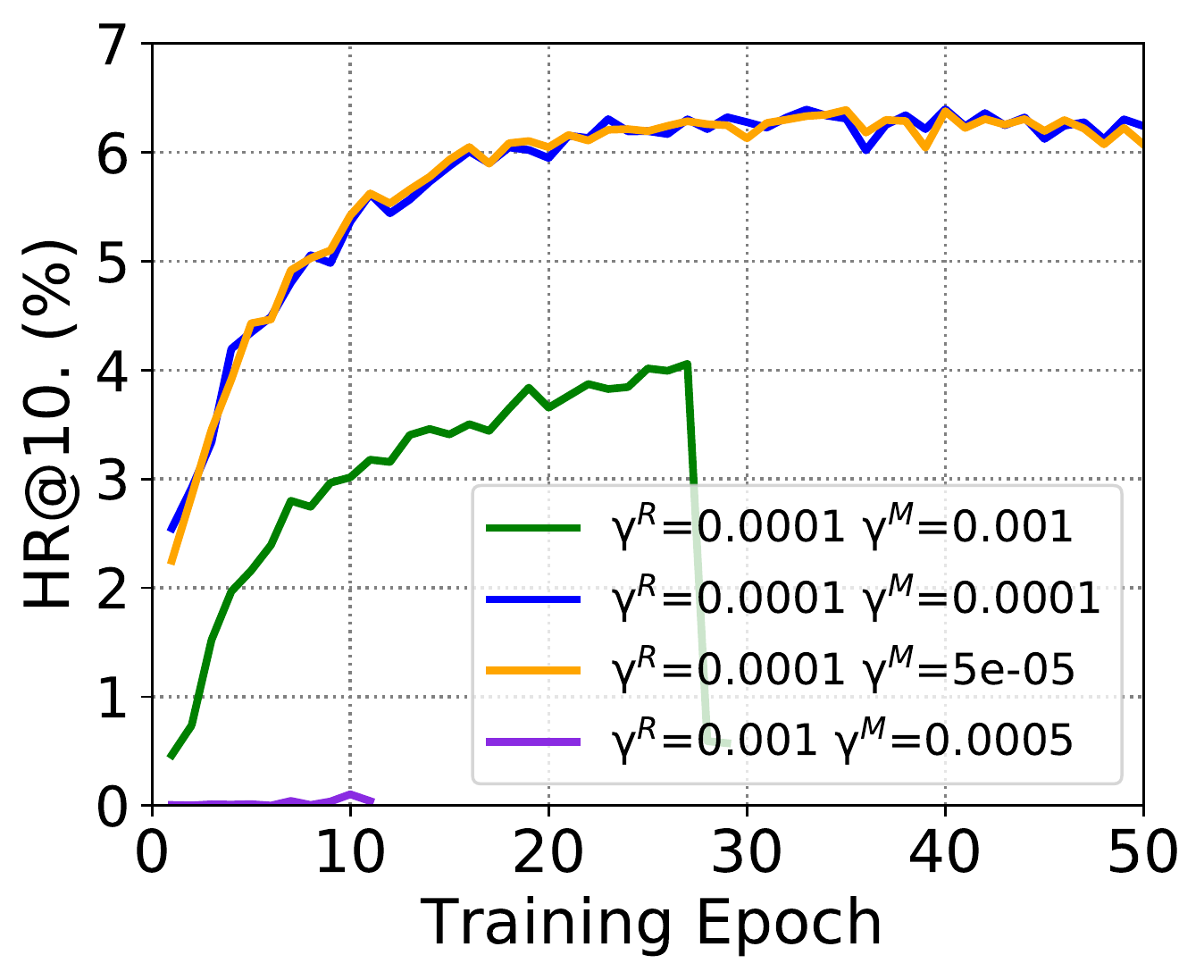}
       }
    \caption{Training collapse (on the validation set) with different learning rates $\gamma$. $M$ and $R$ denote learning rate for ME and the remaining modules, respectively.}
    \label{apx:tb:seaching learning rate}
    \end{figure}

\textbf{Model collapse.}
Unlike IDRec, we find a very surprising phenomenon that
training MoRec without proper hyper-parameters (mainly the learning rate $\gamma$) can easily lead to model collapse. As shown in Figure~\ref{apx:tb:seaching learning rate}, the performance of MoRec on MIND drops drastically from 16\% to 0 when $\gamma^M$ and $\gamma^R$ are equal to 0.0001. Even worse, MoRec becomes collapsed from the beginning when  $\gamma^M=0.0001$ and $\gamma^R=0.001$. Similarly, MoRec also have this problem when making image recommendation on HM. However, by carefully searching hyper-parameters, we find that MoRec  can usually be trained well with a proper $\gamma$. It is worth noting that it is sometimes necessary to set different $\gamma$ for item ME and other modules. This may be because item ME has been pre-trained on NLP and CV datasets before, and its learning stride may be different from other modules trained from scratch. By contrast, IDRec did not collapse even with many different $\gamma$.
To the best of our knowledge, our findings here have not been reported in the literature.

\begin{table}[t]
         \caption{Comparison of HR@10 (\%) w/ and w/o extra pre-training.
			`Improv.' means the relative improvement of w/ extra pre-training compared to w/o extra pre-training.
         }
         \label{tb:sendstagepretraining}
         \begin{center}
         \scalebox{0.85}{
   \begin{tabular}{ cc| ccc ccc}
   \toprule
   \multirow{2}{*}{Dataset} &\multirow{2}{*}{ME}&\multicolumn{3}{c}{TS} &\multicolumn{3}{c}{E2E} \\
   \cmidrule(lr){3-5}\cmidrule(lr){6-8}
   & &w/o &w/  &Improv.&w/o &w/ &Improv.\\
   \midrule
   MIND&$\text{BERT}_{\text{base}}$
   &13.93 &\textbf{14.68} &+5.38\% &18.23 &\textbf{18.63} &+2.19\%\\
   HM&$\text{MAE}_{\text{base}}$
    &2.50 &\textbf{2.79} &+11.60\% &7.03 &\textbf{7.07} &+0.57\%\\
   Bili &$\text{MAE}_{\text{base}}$
   &0.57 &0.57 &0.00\%  &\textbf{3.18}  &3.17 &-0.31\%\\
   \bottomrule
   \end{tabular}}
         \end{center}
         \vspace{-2mm}
      \end{table}

   \section{Related Work}\label{section:Related Work}
   \textbf{ID-based recommender systems (IDRec).}
   In the existing recommendation literature, there are countless models built entirely on user/item ID, from early item-to-item collaborative filtering~\citep{linden2003amazon}, shallow factorization models~\citep{koren2009matrix,rendle2010factorization}, to deep neural models~\citep{he2017neural,hidasi2015session}.
    They can be roughly divided into two categories: non-sequential models (NSM) and sequential neural models (SRM). NSM further includes various recall (e.g., DSSM, and YouTube DNN~\citep{covington2016deep}) and CTR models (e.g., DeepFM~\citep{guo2017deepfm}, wide \& Deep~\citep{cheng2016wide}, and Deep Crossing~\citep{shan2016deep}).
    These models typically take a user-item pair as input along with some additional features and predict matching scores between users and items. In contrast, a typical SRM takes a sequence of user-item interactions as input and generates the probability of the next interaction.
   The most representative SRM includes GRU4Rec~\citep{hidasi2015session}, NextItNet~\citep{yuan2019simple,yuan2021one}, SR-GNN~\citep{wu2019session},
   SASRec~\citep{kang2018self} \& BERT4Rec~\citep{sun2019bert4rec} with RNN, CNN, GNN, Transformer \&  BERT as the backbone, respectively, among which SASRec often performs the best in  literature~\citep{yuan2022tenrec,fischer2021comparison,zhang2022dynamic}.
   
   \textbf{Modality-based recommender systems (MoRec).}
   MoRec  focus on modeling the modality content features of items, such as text~\citep{wu2020mind}, images~\citep{mcauley2015image}, videos~\citep{deldjoo2016content}, audio~\citep{van2013deep} and text-image multimodal pairs~\citep{wu2021mm}. Previous work tended to adopt the two-stage (TS) mechanism by first pre-extracting item modality features from ME and then incorporating these fixed features into the recommendation model~\citep{mcauley2015image,he2016vbpr,he2016ups,shan2016deep,lee2017large,tang2019adversarial,wei2019mmgcn}.  What's more, most of these work mainly use modality as side features and IDs as the main features.
   E2E-based MoRec is not popular until recently for several reasons: (1) the TS mechanism is architecturally very flexible for industrial applications and requires much lower compute and training cost;
   (2) there were few high-quality public datasets with original item modalities; (3) ME developed in past literature (e.g., word embedding) is not expressive enough even with E2E training.
In the past two years, some works have begun to explore E2E-based MoRec, however, most of them focus on text recommendation~\citep{wu2021empowering,shin2021one4all,yu2021tiny,yang2022gram,xiao2022training,hou2022towards}.
   A recent preprint~\citep{elsayed2022end} introduced ResNet as ME for fashion-based recommendation but had to rely on  ID features for competitive accuracy. To the best of our knowledge, none of these existing peer-reviewed literature  provides an explicit and comprehensive comparative study of MoRec and its corresponding IDRec counterpart in a fair experimental setting (e.g., making sure they use the same backbone for comparison), especially in the non cold-start or even warm-start settings.
   

   \section{Conclusion and Future Works}
In this paper, we investigated an ambitious but under-explored question, whether MoRec has the opportunity to end the dominance of IDRec. Obviously, this problem cannot be completely answered in one paper, and requires more study and efforts from the RS and even the NLP and CV communities.
   Yet, one major finding here  is that with the SOTA and E2E-trained ME, \textit{modern} MoRec could already perform on par or better than IDRec with the typical recommendation architecture (i.e., Transformer backbone) even in the non cold-start item recommendation setting.
   Moreover,  MoRec can largely benefit from the technical advances in the NLP and CV fields, which implies that it has larger room for accuracy improvements in the future. Given this, we believe our research is meaningful and would potentially inspire more studies on E2E-based MoRec, for example, developing more powerful recommendation architectures (particular for CTR\footnote{In fact, we notice that the NLP/CV communities are formulating most tasks into sequence learning problem with Transformer as the backbone~\citep{reed2022generalist,raffel2020exploring}, e.g., GPT-3 and pixelGPT~\citep{chen2020generative}. It will be interesting to see whether complex CTR models with various user/item features can be formulated in a similar fashion (the way MoRec is powerful).} prediction tasks), more expressive \& generalized item encoders, better item \& user fusion strategies and more effective optimizations to reduce the compute \& memory costs and the longer training time.
   We also envision that in the long run the prevailing paradigm of RS may have a chance to shift from IDRec to MoRec when raw modality features are available. 
   
  As mentioned above, this study is only a preliminary of MoRec and 
  has several limitations: (1) we  considered RS scenarios with only text and vision, whereas MoRec's behaviors with other modalities, e.g., voice and video, remain unknown; (2) we consider only single-modal item encoders, while the behaviors of multimodal MoRec are unknown; 
  (3) we  considered only a very basic approach to fusing ME into recommendation models, thereby MoRec may achieve  sub-optimal performance;
  (4) our observations were made on  three medium-sized dataset, and
  it remains unknown whether the key findings  hold if we scale up training data to 100$\times$ or 1000$\times$ as in real industrial systems.

\begin{acks}
This work is supported by the Research Center for Industries of the Future (No.WU2022C030) and the Key Research Project of Zhejiang Lab (No.2022PG0AC02).
\end{acks}

\appendix
\section{Appendix}      
\label{ cold-sc}
\subsection{MoRec \emph{vs} IDRec on cold-start settings} 
\label{morecvsidrec}

         \vspace{-2mm}
\begin{table}[htbp]
         \caption{HR@10 (\%) of IDRec and MoRec for cold and new item recommendation. $m^{cold}$ and $m^{new}$ denote the number of cold items and new items, respectively. All results are evaluated based on the SASRec architecture.
         }
         \vspace{-2mm}
         \label{tb:cold_new_item_recommendation}
         \begin{center}
         \scalebox{0.85}{
    \begin{tabular}{ cc|ccc|ccc }
   \toprule
   Dataset&ME&$m^{cold}$ & IDRec &MoRec &$m^{new}$  &IDRec &MoRec\\
   \midrule
   MIND &$\text{BERT}_{\text{base}}$ &32K &0.0036 &\textbf{3.0637}  &13K &0.0125  &\textbf{0.5899} \\
   HM &Swin-B &37K &0.3744  &\textbf{1.0965}   &14K&0.0115  &\textbf{0.6846}\\
   Bili &Swin-B &39K &0.3551 &\textbf{0.6400}&5K  &0.0078 &\textbf{0.0832}\\
   \bottomrule
   \end{tabular}}
         \end{center}
         \vspace{-2mm}
      \end{table}


    MoRec is a natural fit for cold item recommendation as their ME is specifically developed to model the raw modality features of an item,  whether it is cold or not. To validate this, we evaluate IDRec and MoRec in  two scenarios, i.e., COLD item setting and NEW item setting. Specifically, we
  counted the interactions of all items in the training set and regarded those that appeared less than 10 times as cold items.  We found that the number of cold items were very small in our original testing test. So we performed dataset crawling  again for one month and then selected user sequences (from this new dataset) that contained these cold items (as cold item setting) and  items that did not appear in the training set (as new item setting).
    We report the results in Table~\ref{tb:cold_new_item_recommendation}. As expected, MoRec consistently and substantially improve IDRec on all three datasets for both text and vision modalities in both cold and new settings.
    The superiority of MoRec comes from the powerful representations of ME which were pre-trained on large-scale text and image datasets beforehand.

\bibliographystyle{ACM-Reference-Format}
\balance
\bibliography{sample-base.bib}

\end{document}